# Relaxation dynamics and ionic conductivity in a fragile plastic crystal

Th. Bauer,[1] M. Köhler,[1] P. Lunkenheimer,[1,a] A. Loidl,[1] and C. A. Angell[2]

[1]*Experimental Physics V, Center for Electronic Correlations and Magnetism, University of Augsburg, Universitätsstr. 2, 86135 Augsburg, Germany*
[2]*Department of Chemistry and Biochemistry, Arizona State University, P.O. Box 871604, Tempe, Arizona 85287, USA*

We report a thorough characterization of the dielectric relaxation behavior and the ionic conductivity in the plastic-crystalline mixture of 60% succinonitrile and 40% glutaronitrile. The plastic phase can be easily supercooled and the relaxational behavior is investigated by broadband dielectric spectroscopy in the liquid, plastic crystalline, and glassy crystal phases. The very pronounced $\alpha$-relaxation found in the spectra is characterized in detail. From the temperature dependence of the $\alpha$-relaxation time, a fragility parameter of 62 was determined making this material one of the most fragile plastic-crystalline glass formers. A well-pronounced secondary and faint indications for a third relaxation process were found, the latter most likely being of Johari-Goldstein type. In addition, relatively strong conductivity contributions were detected in the spectra exhibiting the typical features of ionic charge transport.

## I. INTRODUCTION

Plastic crystals (PCs) exhibit glassy freezing of the orientational degrees of freedom of the molecules, while their centers of gravity are well ordered on a crystalline lattice. They show many of the unusual and only poorly understood hallmark features of glassy dynamics like non-exponential relaxation and non-Arrhenius behavior of the molecular relaxation times.[1,2] Thus, they are considered as model systems for canonical glass formers and their study can help elucidating the so far mysterious glass-transition phenomenon. However, PCs do not cover the whole range of the parameter space covered by canonical glass formers. Especially, nearly all PCs can be classified as "strong" glass formers within the strong/fragile classification scheme[3] that has proven to correlate with many key properties of glassy matter.[4,5] While fragile glass formers are characterized by pronounced deviations of their temperature-dependent $\alpha$-relaxation times $\tau_\alpha$ from Arrhenius behavior, these deviations are relatively small in most PCs.[2,6,7,8] Another typical property of PCs is the finding of no or only rather weak secondary relaxation processes.[2,9,10,11] Secondary or $\beta$-relaxations seem to be a relatively universal feature of canonical glass formers.[12,13] Only a minority of them exhibits a so-called excess wing instead,[14,15,16] which, however, also can be ascribed to a secondary relaxation.[17]

An exception to this typical behavior of PCs is provided by Freon112, which stands out by being rather fragile with a fragility index $m \approx 68$ and by having a very strong $\beta$-relaxation.[18] This finding was ascribed to the fact that the Freon112 molecules can assume two different conformations, *trans* and *gauche*: Based on the inverse proportionality of effective energy barrier and configurational entropy deduced within the Adam-Gibbs theory,[19] the fragility of a glass former can be assumed to be linked to the form of the potential energy landscape in configuration space.[20] Within this framework a higher density of minima should lead to higher fragility. For PCs, it seems reasonable that the lattice symmetry leads to a reduced density of energy minima, which explains their relatively low fragility. Thus, in Ref. 18 the observed high fragility of Freon112 was rationalized by assuming that its *trans-gauche* disorder leads to a higher density of minima in the potential energy landscape. In addition, also the strong $\beta$-relaxation in this PC was related to its conformational disorder, namely to the fact that its *trans* conformer is nonpolar and thus Freon112 may be regarded as being composed of polar *gauche* molecules "dissolved" in a nonpolar medium.[18]

Aside from Freon112, to our knowledge only one PC was reported in literature to also exhibit relatively high fragility, namely the mixture of 60% succinonitrile and 40% glutaronitrile (60SN-40GN).[6,7] In Refs. 6 and 7 the temperature dependence of the $\alpha$-relaxation times was shown, however, in a relatively limited temperature range only. In addition, no dielectric spectra were reported until now and nothing is known about the other parameters of the $\alpha$-relaxation and about the possible occurrence of secondary relaxations. Thus for the present work we have performed a thorough characterization of this interesting PC. In addition, we provide the phase diagram of the SN-GN mixtures.

## II. EXPERIMENTAL DETAILS

The sample materials were purchased from Arcos Organics and measured without further purification. The specified purities for succinonitrile and glutaronitrile were $\geq 99.0\,\%$. The mixtures were prepared in a glovebox under Ar atmosphere. Liquid glutaronitrile was put into succinonitrile, melted in a water bath, under heavy stirring. The concentrations are specified in mol%. To check for phase transitions and glass anomalies, the sample materials were characterized by differential scanning calorimetry (DSC) using typical heating rates of 10K/min.

To record the real and imaginary parts of the dielectric permittivity in a frequency range from $10^{-2}$ to $3\times10^9$ Hz, two different techniques were combined: At low frequencies, $\nu < 3$ MHz, a frequency-response analyzer (Novocontrol $\alpha$-analyzer) was used. At frequencies $\nu > 1$ MHz the impedance analyzer HP4291 was employed for a reflectometric technique, where the sample capacitor is mounted at the end of a 7 mm coaxial line [21]. For both techniques, polished parallel-plate capacitors were used and filled with the liquid sample material. The plates were kept at distance using glass-fiber spacers with typical diameters of 100 µm. Cooling and heating of the samples was achieved by a nitrogen gas-heating system (Novocontrol Quatro). Moderate cooling rates of 0.5 K/min were used. For further experimental details the reader is referred to refs. [16], [21], and [22].

---

[a]Electronic mail: peter.lunkenheimer@physik.uni-augsburg.de

## III. RESULTS AND DISCUSSION

Figure 1 shows the phase diagram of the mixture of SN and GN as determined by DSC measurements. For this purpose, the samples were quickly cooled and the DSC signal measured under heating. For GN-concentrations $x \leq 10\%$, a plastic phase arises, which, however cannot be supercooled. At $15\% \leq x \leq 95\%$ supercooling is possible, arriving at a so-called glassy-crystal state.[23] At $x \geq 75\%$, there is an increased tendency to assume complete order, which arises during heating at the temperatures indicated by the crosses in Fig. 1. For pure GN ($x = 100\%$) no plastic phase is found but below 253 K the sample is translationally and orientationally ordered.

In Ref. 24 it was proposed that a material should have the ability to form a glassy crystal if $T_b / T_{tr} \geq 2.5$. Here $T_b$ is the boiling point and $T_{tr}$ is the transition temperature to the completely ordered crystalline state. This relation is analogous to a rule for the glassforming ability in canonical glass formers, $T_b / T_m > 2$.[25] In the case of pure SN, $T_b / T_m = 540/327 = 1.65$ and $T_b / T_{tr} = 540/235 = 2.30$, predicting neither the formation of a structural glass nor of a glassy crystal. However, for $x = 15\%$, from the phase diagram $T_{tr} \approx 220$ K can be expected. This leads to $T_b / T_{tr} = 2.45$ and indeed the glassy-crystal state sets in at this concentration in the phase diagram. In addition, the common finding for structural glass formers, $T_g / T_m \approx 2/3$, is upheld when $T_m$ is replaced by $T_{tr}$ ($T_g / T_m = 147/220 = 0.67$). In the present work, we provide a detailed dielectric characterization of the mixture with $x = 40\%$, which is closer to the minimum in $T_m(x)$ and shows negligible tendency to convert into the completely ordered state. Dielectric investigations of additional concentrations will be provided in a forthcoming paper.

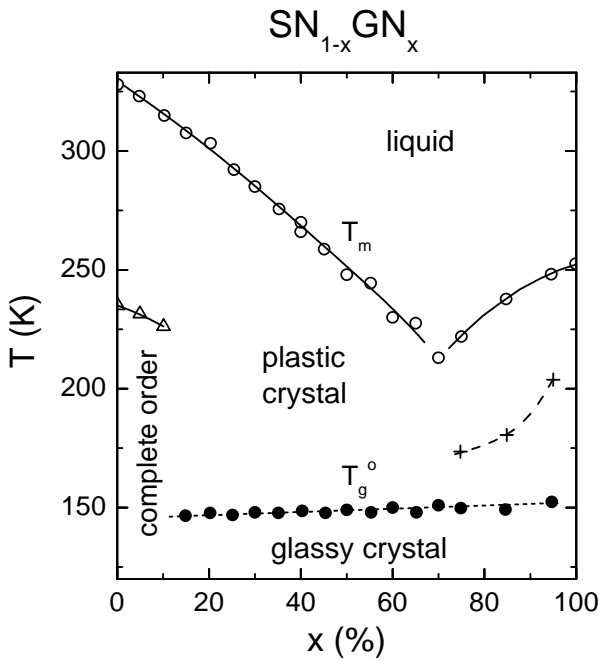

FIG. 1. Phase diagram of the system $SN_{1-x}GN_x$ as determined from DSC measurements under heating. The open and closed circles denote the melting and the orientational-glass temperatures, respectively. The triangles indicate the transitions from the completely ordered (i.e., both translationally and orientationally) to the plastic-crystalline state. The crosses correspond to the occurrence of complete order under heating, occurring at high $x$ values.

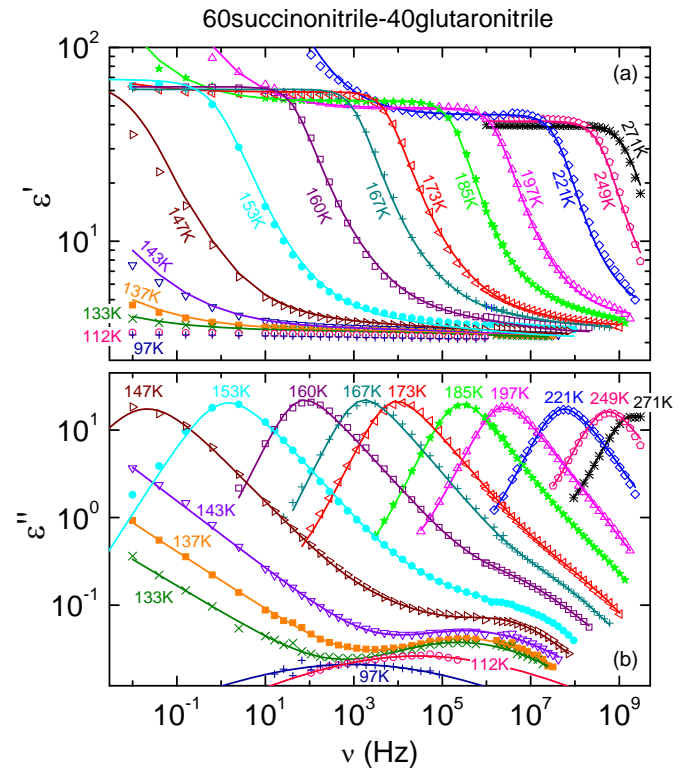

FIG. 2. Frequency dependence of $\varepsilon'$ (a) and $\varepsilon''$ (b) of 60SN-40GN for selected temperatures. The lines are fits with a CD function for $T \geq 197$ K and the sum of a HN and a CC function at $T \leq 185$ K. Except for 147 and 153 K, the parameter $\alpha$ of the HN function was set to zero, corresponding to a CD function. In addition, dc- and ac-conductivity contributions were used in the fits for $T \geq 160$ K. Both the real and imaginary parts of the permittivity were fitted simultaneously.

Figure 2 shows spectra of the dielectric constant $\varepsilon'$ and the loss $\varepsilon''$ of 60SN-40GN measured at various temperatures under cooling. DSC measurements (not shown) revealed a transition of the liquid 60SN-40GN mixture into the plastic-crystalline phase at 250 K. Thus, except for 271 K, all curves in Fig. 2 were taken in that phase. The spectra exhibit the well-known features of $\alpha$-relaxation, namely a step-like decrease of $\varepsilon'(\nu)$ and corresponding peaks in the loss. Their strong shifting towards lower frequencies with decreasing temperature mirrors the glassy freezing of the orientational degrees of freedom. Assuming the often-employed criterion that at the glass transition the relaxation time $\tau$ should be of the order of 100 s, corresponding to a loss-peak frequency of $\nu_p = 1/(2\pi\tau) \approx 10^{-3}$ Hz, already from a first glance at Fig. 2(b) the orientational glass temperature $T_g^o$ in this PC can be estimated to be around 145 K. This agrees with the typical glass anomaly found in the DSC measurements at about 149 K (cf. Fig. 1). Thus, in contrast to pure succinonitrile, which undergoes a transition to complete orientational order at about 235 K (Fig. 1),[26,27] in 60SN-40GN the plastic phase can be easily supercooled, finally arriving in a glassy-crystal state.[23] Aside from the $\alpha$-relaxation, another interesting feature of the spectra in Fig. 2 is the emergence of a clear secondary relaxation peak at low temperatures. As typical for $\beta$-relaxations, at high temperatures it successively merges with the $\alpha$-peak, becoming indiscernible at $T \geq 197$ K. At the lowest temperatures, where the secondary peaks remain unobscured by the $\alpha$-relaxation, they are revealed to be symmetrically broadened and can be well fitted by the Cole-Cole (CC) function,[28]

$$\varepsilon^* = \varepsilon' - i\varepsilon'' = \varepsilon_{\infty,CC} + \frac{\Delta\varepsilon_{CC}}{1 + (i\omega\tau_{CC})^{1-\alpha_{CC}}} \quad (1)$$

as commonly found for secondary relaxations [lines through the $\varepsilon''(\nu)$ curves at 97 and 112 K in Fig. 2(b)].

Following common practice, for the sake of clarity in Fig. 2(b) the contributions from impurity-induced charge transport, leading to a divergence of $\varepsilon''(\nu)$ at frequencies below the left wing of the $\alpha$-peaks, are not shown. However, as these contributions in 60SN-40GN show some peculiarities, in Fig. 3 we treat them in more detail. Conductivity effects are most pronounced at the higher temperatures. As a typical example, the left inset of Fig. 3 shows the loss at 221 K including the mentioned low-frequency increase due to charge transport. Interestingly, the usual $\varepsilon'' \sim 1/\nu$ behavior, arising from dc conductivity via the relation $\varepsilon'' = \sigma'/(2\pi\nu\varepsilon_0)$ (with $\varepsilon_0$ the permittivity of vacuum), is not able to describe the curves: The line in this inset is the sum of this dc contribution and a Cole-Davidson (CD) law,[29]

$$\varepsilon^* = \varepsilon_{\infty,CD} + \frac{\Delta\varepsilon_{CD}}{(1 + i2\pi\nu\tau_{CD})^{\beta_{CD}}} \quad (2)$$

as often employed for the description of the $\alpha$-relaxation. Obviously, the experimental data show a more shallow minimum than expected, which we also found for other temperatures in the plastic phase. In the lower inset of Fig. 3 the same data set is shown in conductivity representation, revealing additional ac conductivity, namely a power-law increase $\nu^s$ with exponent $s < 1$ as possible origin of these deviations (dash-dotted line). Thus overall we arrive at

$$\sigma' = \sigma_{dc} + \sigma_0 \nu^s \quad (3),$$

where $\sigma_{dc}$ denotes the dc conductivity and $\sigma_0$ is a prefactor. Such a behavior is typical for hopping charge transport as found in many kinds of disordered matter and in ionic conductors.[30,31,32] Towards low frequencies, $\nu < 10$ Hz, the conductivity $\sigma'(\nu)$ for 221 K starts to decrease again, reaching values below the dc plateau. This points to blocking-electrode effects, again typical for ionic conductors.[33] Indeed, succinonitrile has been shown to provide the basis for high conductivity solid phases when serving as a solvent for salts that can dissolve in the rotator phase lattice[27,34] (though a contribution from a residual liquid saline solution phase at the grain boundaries has not been rigorously excluded). As shown in the main frame of Fig. 3, the complete $\sigma'(\nu)$ curves (and thus also the $\varepsilon''(\nu)$ spectra) at high temperatures can be described by the sum of dc and ac conductivity for the ionic charge transport and a CD function for the relaxation, which in the $\sigma'$ representation shows up as a shoulder.

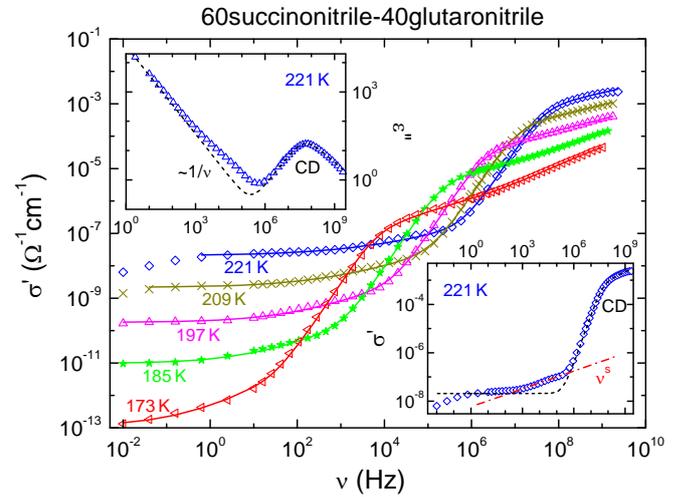

FIG. 3. Frequency dependence of the conductivity of 60SN-40GN for selected temperatures. The lines are fits with the sum of dc and ac conductivity ($\sigma' \sim \nu^s$), added to the conductivity-equivalent of the CD function [calculated from eq. (2) via $\sigma' = 2\pi\nu\varepsilon''\varepsilon_0$]. The same fits in loss representation are shown in Fig. 2(b). The left inset shows the dielectric loss for 221 K, including the conductivity-induced divergence at low frequencies. The dashed line, calculated assuming a CD function and dc conductivity cannot describe the results. The right inset shows the same 221 K data and the calculated curve in conductivity representation. The dash-dotted line indicates a $\nu^s$ increase, typical for ionic hopping charge transport.

However, one should be aware that, instead of an $\nu^s$ contribution, also other descriptions of the excess intensity revealed in the left inset of Fig. 3 may be possible, e.g., assuming an additional broad relaxation peak at frequencies below that of the $\alpha$-peak. Such a relaxation could be due to transitions between different conformers of the molecules. Indeed, for the plastic phase of pure succinonitrile, neutron scattering experiments suggest a *trans-gauche* transition at a rate slower than that of the $\alpha$-relaxation.[35] However, nothing is known about the characteristic times of this or other conformational transitions in the SN-GN mixture and at the temperatures investigated in the present work.

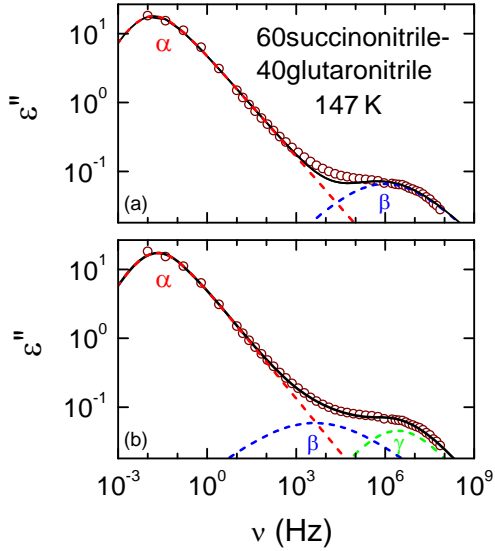

FIG. 4. Frequency dependence of the loss at 147 K. (a) and (b) show two alternative scenarios for the description of the loss spectra at low temperatures: The solid lines show fits with the sum of a CD and a CC function (a) or with a CD and two CC functions (b). The dashed lines show the different contributions to the fits.

The complex permittivity curves at $T \geq 160$ K in Fig. 2 were thus fitted using a CD function including dc *and* ac conductivity (lines in Fig. 2). Here also the ac contribution in $\sigma''$, following from eq. (3) via the Kramer-Kronig relation, was taken into account, namely $\sigma'' = \tan(s\pi/2)\,\sigma_0 \nu^s$ leading to $\varepsilon' = \sigma'/(2\pi\nu\varepsilon_0) \sim \nu^{s-1}$.[30] It explains the additional low-frequency increase of $\varepsilon'(\nu)$ beyond the upper plateau value of the relaxation step [Fig. 2(a); not completely shown]. At $T \leq 185$ K the emerging secondary relaxation was described by using an additional Cole-Cole function in the fits. At $T \leq 153$ K the left flanks of the $\alpha$-peaks become rather shallow, which cannot be taken into account by the CD function and instead a Havriliak-Negami (HN) function[36] employing an additional width parameter $\alpha_{HN}$ had to be used. At $T \leq 157$ K, the transition region between $\alpha$- and secondary relaxation is too smeared-out to achieve good fits with two relaxation functions only. This is demonstrated in Fig. 4(a) for the 147 K curve. When assuming a third relaxation, the experimental data in this region can be well fitted [Fig. 4(b)]. However, at the lowest temperatures, $T \leq 145$ K, no clear statement on the presence of this third relaxation can be made because the $\alpha$-peak has shifted out of the frequency window and a single power law is sufficient to describe the spectra at frequencies below the left wing of the well-pronounced secondary peak. Thus the presence of a third relaxation as indicated in Fig. 4(b) is quite speculative and the obtained fitting parameters have large uncertainty. If such a relaxation indeed exists, it may be termed $\beta$-relaxation and the well pronounced secondary relaxation could be designated as $\gamma$-relaxation. In the following we use this nomenclature. Following the principle to minimize the number of fitting parameters as far as possible, we do not use fits with three peak functions in the temperature region above 157 K as here two relaxations are sufficient to describe the data.

The temperature-dependent relaxation strengths and width parameters of $\alpha$- and $\gamma$-relaxation, resulting from the fits in Fig. 2, are plotted in Fig. 5. As revealed by Fig. 5(a), the $\alpha$-relaxation strength $\Delta\varepsilon_\alpha$ roughly follows a Curie-Weiss law (line), $\Delta\varepsilon_\alpha \sim 1/(T-T_{CW})$. The resulting small Curie-Weiss temperature of $T_{CW} = 3.3$ K implies only slight deviations from the Curie behavior expected within Onsager theory. As the $\alpha$-peaks partly were fitted by a HN function instead of a CD function, in Fig. 5(c) we use the high-frequency power-law exponent of the loss peaks, $\beta_\alpha$, as a measure of the peak width [for the CD function, eq. (2), $\beta_\alpha = \beta_{CD}$ and for the HN function $\beta_\alpha = \beta_{HN}(1-\alpha_{HN})$]. After an initial increase, $\beta_\alpha(T)$ tends to saturate at a value significantly smaller than unity [Fig. 5(c)]. Such a behavior is rather common in glass forming liquids[16,37,38] and also found in other PCs.[2] The amplitude of the $\gamma$-relaxation $\Delta\varepsilon_\gamma$ increases [Fig. 5(b)] and its width parameter $\alpha_\gamma$ [$= \alpha_{CC}$ in eq. (1)] decreases [Fig. 5(d)], i.e. the peaks become narrower, with increasing temperature, a behavior often found for secondary relaxations.[13,38] Finally, the relaxation strength of the suspected $\beta$-relaxation varies between 0.35 and 0.48 and its width parameter covers a range of 0.48 - 0.65 (not shown). As information on this relaxation could be obtained in the limited region of 147 - 157 K only, no significant statement on the temperature-dependent trends of these parameters can be made.

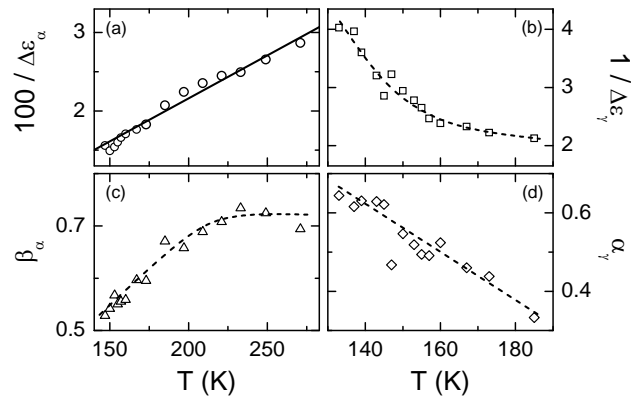

FIG. 5. Inverse relaxation strengths (a,b) and width parameters (c,d) of the $\alpha$- and $\gamma$-relaxation obtained from the fits shown in Fig. 2. The solid line in (a) is a linear fit demonstrating approximate Curie-Weiss behavior. The dashed lines are drawn to guide the eyes.

The most significant parameters of relaxational processes in glass formers are the temperature-dependent relaxation times. Their values, as resulting from the fits shown in Fig. 2, are plotted in Fig. 6 in Arrhenius representation.[39] The $\alpha$-relaxation times $\tau_\alpha$ reveal clear deviations from Arrhenius behavior. A good empirical parameterization of $\tau_\alpha(T)$ can be achieved by the commonly employed modified Vogel-Fulcher-Tammann (VFT) law,[3,40]

$$\tau_\alpha = \tau_0 \exp\left[\frac{DT_{VF}}{T - T_{VF}}\right] \qquad (4),$$

with $T_{VF}$ the Vogel-Fulcher temperature and $D$ the strength parameter[3] (solid line in Fig. 6). Also alternative descriptions are possible, e.g., similar to Freon112,[41] the data can well be fitted by the recently proposed formula by Mauro et al.[42] (dotted line). Using the condition $\tau_\alpha(T_g) \approx 100$ s, for both fit curves an extrapolation yields an orientational-glass temperature $T_g^o$ of 144 K.

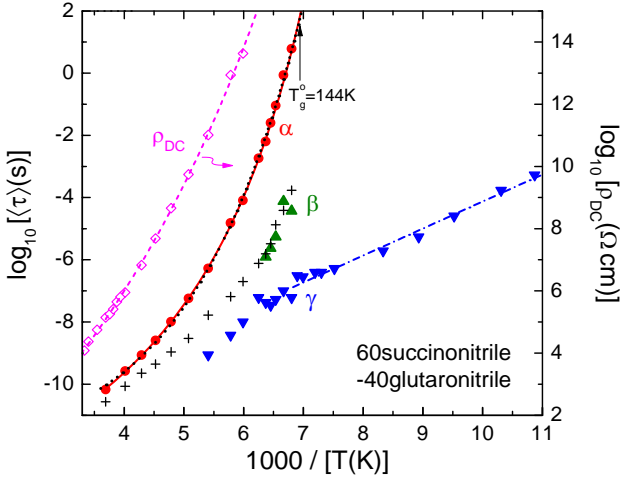

FIG. 6. Average relaxation times[39] (closed symbols, left axis) and dc resistivity (open diamonds, right axis) determined from the fits shown in Fig. 2. The solid line is a fit of $\tau_\alpha(T)$ with a VFT function, eq. (4) ($\tau_0 = 1.8\times10^{-14}$ s, $T_{VF} = 106$ K, and $D = 12.8$). The dotted line is a fit using the formula by Mauro et al.[42] ($\tau_0 = 1.2\times10^{-12}$ s, $K = 110$ K, $C = 416$ K). The dash-dotted line is a fit of the $\beta$-relaxation time at $T < T_g$ assuming an Arrhenius law ($\tau_0 = 1.6\times10^{-13}$ s, $E = 0.17$ eV). The crosses show the prediction for $\tau_{JG}$ calculated from eq. (5).[48] The dashed line is a fit of the dc resistivity with a VFT-like function, $\rho_{dc} = \rho_0 \exp[D_\rho T_{VF}^\rho / (T - T_{VF}^\rho)]$ leading to $\rho_0 = 4.5\times10^{-3}$ s, $T_{VF}^\rho = 77.9$ K, and $D_\rho = 42.5$.

The relatively small strength parameter $D = 12.8$, obtained from the VFT fit, shows that plastic crystalline 60SN-40GN indeed stands out among most other PCs by exhibiting fragile characteristics of its relaxation dynamics. This becomes especially obvious in Fig. 7, which provides a $T_g$-scaled Arrhenius plot[43] of the $\alpha$-relaxation times of various PCs.[2,18,44] Only 60SN-40GN and Freon112 exhibit pronounced fragile characteristics. From the slopes of these curves at $T_g$ the fragility parameter $m$ can be determined.[45] From Fig. 7, we find $m = 62$, characterizing 60SN-40GN as a fragile glass former. Its fragility is comparable to $m = 68$ in Freon112[18] and in Fig. 7 the $\tau_\alpha(T)$ curves of both substances nearly match. An empirical correlation between the fragility and the width parameter of the $\alpha$-relaxation, $\beta_{KWW}$, was found to hold for numerous glass formers.[5] $\beta_{KWW}$ is the stretching parameter of the Kohlrausch-Williams-Watts function[46] and can be calculated from $\beta_{CD}$.[47] The obtained $m = 62$, $\beta_{KWW} \approx 0.60$ for 60SN-40GN matches well the reported correlation of both quantities.[5]

The relaxation times of the suspected $\beta$-relaxation and of the well-pronounced secondary relaxation, denoted as $\gamma$-relaxation are also included in Fig. 6. Below about 155 K, the $\gamma$-relaxation times follow Arrhenius behavior with an energy barrier of 0.17 eV. At higher temperatures the fit results indicate a transition to stronger temperature dependence of $\tau_\gamma(T)$. However, as this is the region of merging with the $\alpha$-relaxation and, in addition, an influence of the $\beta$-relaxation cannot be excluded, this observation is of limited significance. The $\beta$-relaxation times could be determined in a rather restricted temperature region only and thus no clear statement on the functional form of its temperature dependence can be made. However, a description with an Arrhenius law seems unreasonable because it becomes obvious in Fig. 6 that this would lead to an unreasonably small attempt frequency $\tau_0$ (the value of $\tau_\beta$ for $1000/T \to 0$).

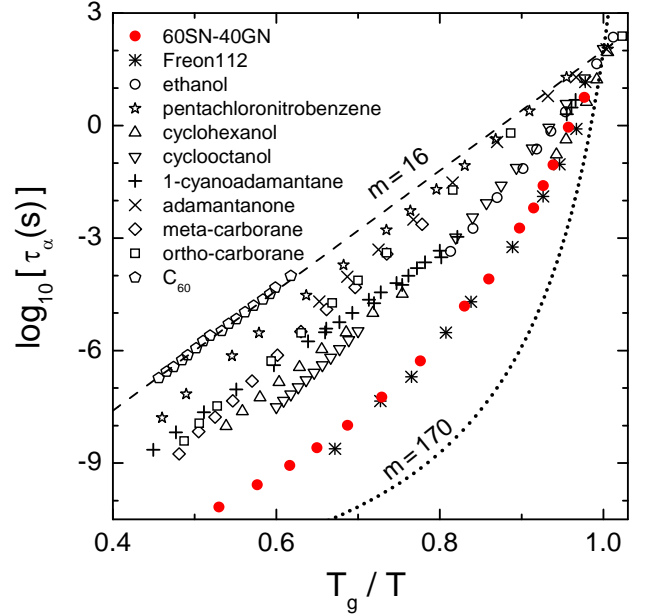

FIG. 7. $T_g$-scaled Arrhenius plot of the $\alpha$-relaxation times of various PCs.[2,18,44] The results of the present work are presented as closed circles. The dashed and dotted lines show the behavior for minimal and very high fragilities.

A considerable fraction of the recent literature on glassy dynamics deals with the investigation of the so-called Johari-Goldstein[12] (JG) relaxations, which are thought to be inherent to the glassy state of matter but whose microscopic origin still is unclarified. This type of secondary relaxation should not be confused with other, less universal relaxations, e.g., caused by intramolecular motions, which are of minor interest for the understanding of the glass transition. To identify "genuine" JG relaxations, in Ref. 48 a criterion was proposed, which relates the JG-relaxation time $\tau_{JG}$ to the parameters of the $\alpha$-relaxation, namely

$$\tau_{JG} \approx (t_c)^{1-\beta_\alpha} (\tau_\alpha)^{\beta_\alpha} \qquad (5),$$

where $t_c \approx 2$ ps. $\tau_{KWW}$ and $\beta_{KWW}$ are parameters of the Kohlrausch-Williams-Watts function,[46] whose Fourier transform often is employed to fit the $\alpha$-relaxation. Both quantities also can be calculated from the corresponding CD parameters.[47] The crosses in Fig. 6 show that $\tau_{JG}(T)$, calculated in this way, agrees surprisingly well with the $\beta$-relaxation times obtained from the fits in Fig. 2. Thus, within the framework promoted in Ref. 48 one may identify this relaxation with the genuine JG relaxation in this material. The situation seems to be similar to that found in several glass-forming liquids where the JG relaxation peak was reported to be hidden between $\alpha$- and $\gamma$-peak in the standard dielectric spectra and its presence could only be revealed under elevated pressure[49] or by mixing the material with an apolar host glass former.[50] However, as mentioned above, the deduced $\tau_\beta$ values shown in Fig. 6 have high uncertainty as no clear shoulders or peaks are observed in the loss spectra. Thus the designation of this relaxation as genuine JG process can be tentative only and needs to be corroborated by further investigations.

Finally, Fig. 6 also shows the temperature dependence of the dc resistivity $\rho_{dc} = 1/\sigma_{dc}$ as determined from the fits of the dielectric spectra (Figs. 2 and 3). The same axis scaling (i.e., decades/cm) was chosen as for the relaxation-time data. In glass-forming dipolar liquids, the temperature dependence of the dc resistivity, which is governed by charge transport of ionic impurities, often follows that of the $\alpha$-relaxation time. This can be ascribed to the approximate validity of the Debye-Stokes-Einstein (DSE) relation $\tau_\alpha \sim 1/\sigma_{dc}$ although one should note that also deviations from the DSE relation are often discussed.[51] In any case, it is clear that for PCs the DSE should not be valid because rotational and translational motions are completely decoupled. Indeed $\rho_{dc}$ and $\tau_\alpha$ in Fig. 6 show distinctly different behavior, $\rho_{dc}(T)$ revealing much weaker deviations from Arrhenius behavior than $\tau_\alpha(T)$. It can be fitted by a Vogel-Fulcher like formula, $\rho_{dc} = \rho_0 \exp[D_\rho T_{VF}^\rho / (T - T_{VF}^\rho)]$ but with clearly different parameters $D_\rho = 42.5$ and $T_{VF}^\rho = 77.9$ K than the $\alpha$-relaxation time ($D = 12.8$ and $T_{VF} = 106$ K). The fact that a VFT-like law and not a simple thermally activated Arrhenius behavior fits the $\rho_{dc}(T)$ curve is indicative of a separate "glass transition" of the mobile-ion subsystem.[6] The rather low fragility of this subsystem is in line with the findings in other glassy ionic conductors.[6] In Ref. 6 it was ascribed to the confinement-generated low rate of entropy production above the ionic glass transition in the framework of an Adam-Gibbs[19] approach. In canonical glass formers, in which the ionic motions are fully coupled to the motions of the particles controlling the structural relaxation, an ionic conductivity at the glass transition of about $10^{15}$ $\Omega$cm is expected.[6,52] In the present decoupled case, we can use this value to define a glass temperature $T_g^{ion}$ of the mobile-ion subsystem. Extrapolating the $\rho(T)$ curve in Fig. 6 to this value leads to $T_g^{ion} \approx 161$ K. It should be noted that part of the $\rho(T)$ data-points shown in Fig. 6 were taken above the melting temperature of about 250 K. Interestingly, the ionic charge transport is not affected by the transition from the liquid to the plastic crystalline state, i.e. the disorder introduced by the reorientational motion is as favorable for the ionic conduction as the complete disorder (translational *and* reorientational) that is present in the liquid.

## IV. SUMMARY AND CONCLUSIONS

In the present work, aside from the determination of the phase diagram of SN-GN mixtures, we have provided a thorough characterization of the relaxation properties and conductivity of plastic crystalline 60SN-40GN. The orientational disorder can be easily supercooled ($T_g^o \approx 144$ K), leading to the typical relaxational properties as also found in supercooled liquids. Our measurements revealed a well-pronounced secondary relaxation in 60SN-40GN. In addition, indications for a further relaxation having characteristic times between those of the former and the $\alpha$-relaxation were found. Based on Ngai's criterion,[48] we tentatively classify it as genuine JG relaxation. The dielectric spectra of 60SN-40GN show a relatively strong ionic-conductivity contribution. The charge carriers can be assumed to arise from ionic impurities, which are almost unavoidable in polar liquids. The temperature-dependent ionic resistivity follows a Vogel-Fulcher like law and a separate glass transition of the mobile-ion subsystem at about 161 K can be assumed.

Maybe the most interesting aspect of this plastic crystalline material is the marked non-Arrhenius temperature-dependence of its $\alpha$-relaxation time and the corresponding, relatively high fragility of $m = 62$. To our knowledge, similarly high fragility was only found in one further PC, Freon112.[18] This finding corroborates the explanation of the high fragility of Freon112, suggested in Ref. 18, assuming a higher density of minima in the potential energy landscape due to its conformational disorder. Both components of the 60SN-40GN mixture are known to exist in different molecular conformations, too.[35,53] In addition, of course there is strong substitutional disorder in this mixture.[54] Thus, a much higher density of states compared to most other PCs can be easily rationalized. However, it should be noted that for Freon112, based on a calorimetry study,[55] its two conformers can be seen to exchange at a rate several decades slower than the $\alpha$-relaxation. A similar scenario also could prevail in the present case and one may ask how this would affect the energy landscape and fragility. Unfortunately, no clear statement is possible about the dynamics of conformational transitions in the mixture investigated in the present work.

Another aspect that may be of relevance for explaining the high fragility of 60SN-40GN and Freon112 is the non-globular shape of their molecules. Most PCs are composed of rather symmetrical molecules and the most strong ones in fact have nearly globular or disk-like shapes (e.g., ortho-carborane with $m = 20$,[56] adamantanone with $m = 19$,[2] fullerene with $m = 16$,[44,57] or thiophene with $m = 16$)[7]. In Ref. 2 it was already speculated that a globular molecular shape could be correlated with fragility. The deviations from Arrhenius behavior in fragile glass formers can be assumed to arise from an increasing cooperativity of the molecular motions leading to an increase of the effective energy barrier at low temperatures. In contrast, in a strong glass former the molecules can be considered to reorient nearly independently from each other. One may speculate that in most PCs the nearly globular shape of their molecules generates less cooperativity making these materials strong glass formers. This is in line with the finding of low cooperativity in highly symmetrical cyanoadamantane reported in Ref. 8. In contrast, a non-globular shape of the molecules, like in 60SN-40GN and Freon112, may favor higher cooperativity, e.g., due to steric hindrance preventing the reorientation of one molecule without a corresponding motion of its neighbors. Indeed for pure succinonitrile, steric hindrance was shown to play an important role.[58] Most PCs are formed of highly symmetrical molecules as it is easy to conceive that this is favorable for the formation of orientational disorder and thus most PCs are strong glass formers. In 60SN-40GN and Freon112 the additional substitutional and/or conformational disorder seems to hamper complete crystallization and enable the formation of plastic phases despite their non-symmetrical molecules, which makes them the most fragile plastic crystals known so far.